\documentclass[showpacs,amssymb,twocolumn,aps,eqsecnum]{revtex4}
\usepackage{amssymb}
\usepackage{amsmath}
\usepackage[pdftex]{graphicx}
\usepackage{epstopdf}
\usepackage{booktabs}
\usepackage[utf8]{inputenc}

\usepackage{color}

\usepackage{hyperref}

\usepackage{natbib}

\usepackage{fancyhdr}
\usepackage{slashed}

\newcommand{\tr}{\mbox{tr}}

\begin{document}

\date{\today}
\title{Classical Boltzmann equation and high-temperature QED}
\author{F. T. Brandt, R. B. Ferreira and  J. F. Thuorst}
\affiliation{Instituto de F\'{\i}sica, Universidade de S\~ao Paulo,
S\~ao Paulo, SP 05508-090, Brazil}


\begin{abstract}
The equivalence between thermal field theory and the Boltzmann transport equation is investigated 
at higher orders in the context of quantum electrodynamics. We compare the contributions obtained from
the collisionless transport equation with the high temperature limit of the one-loop thermal Green's function. Our approach employs
the representation of the thermal Green's functions in terms of forward scattering amplitudes. The general structure of these amplitudes clearly indicates
that the physics described by the leading high temperature limit of quantum electrodynamics can be obtained from 
the Boltzmann transport equation. We also present some explicit examples of this interesting equivalence. 
\end{abstract}

\maketitle

\section{Introduction}

This paper is about hard thermal loop (HTL) amplitudes in finite temperature quantum electrodynamics (QED). This has been the subject of many investigations not only in QED but also in non-Abelian gauge theories and gravity. These HTL amplitudes have loop momenta of the order of the temperature, $T$, which is considered large compared with all the external momenta.  In the case of non-Abelian gauge field theories, it is possible to sum all the amplitudes in terms of a closed form expression for the  effective action \cite{Frenkel:1989br,Taylor:1990ia,Braaten:1990az}. Similarly, in gravity, closed form expressions for the static and the long wavelenght limits of the effective action have been obtained \cite{Brandt:2009ht,Francisco:2013vg}. The main motivation for these investigations is the search for a consistent thermal field theory in the perturbative regime.

It is well known that the thermal effective action in QED has a leading temperature behavior proportional to $T^2$ which arises only from the contribution of the photon polarization tensor. All the other thermal amplitudes are either subleading or they vanish as a consequence of the neutrality of the plasma.
On the other hand, it is also known that the HTL amplitudes can be generated using the Boltzmann transport equation.
This approach has been used in scalar, non-Abelian gauge theories and gravity  \cite{Elze:1989un,Brandt:1994mv,Brandt:1998bp}.

When we adopt the Boltzmann transport equation approach, in the case of QED, we realize that there is ground for considering some more general possibilities for the thermal amplitudes (although restricted to the semiclassical limit). First, since the plasma does not need to be neutral,  amplitudes with an odd number of external photons can be generated. Second, amplitudes which are not simply proportional to a power of $T$ (like the $T^2$ behavior of the two-photon amplitude) are also generated. (This is the key distinction between the non-Abelian and Abelian theories, since in the former, each amplitude generated by the  
Boltzmann transport equation formalism is proportional to the same power of $T$.)
Furthermore, all these contributions are generated from a manifestly gauge invariant effective action, simply because, as we will see, the Boltzmann transport equation formalism yields an effective action which is expressed in terms of the electromagnetic field tensor $F_{\mu\nu}$. 

However, it is not known if the same $n$-photon amplitudes can also be obtained from a first principle approach, like QED at finite temperature.
Our main goal in the present work is to investigate this issue using the imaginary time formalism \cite{kapusta:book89,lebellac:book96,das:book97}. The main motivation of this investigation is to extend the concept of HTL amplitudes including all the contributions which arise from the Boltzmann transport equation approach. In fact, considering that the Boltzmann transport equation formalism is a classical limit, it is important to understand how it relates with the high temperature limit of QED at finite temperature. As a first step toward this goal, we investigate in the present paper the $n$-photon amplitudes in both formalisms.

In the next section, we review the Boltzmann transport  equation approach and derive the effective action and the corresponding thermal amplitudes.  
We obtain an implicit expression for all the thermal amplitudes and compute some explicit examples. 
In Sec. \ref{sec3} we consider the thermal Green's functions in QED at finite temperature. We review how the thermal Green's functions can be expressed in terms of forward scattering amplitudes. Then, using the HTL limit, we argue that the leading behavior of each of these forward scattering amplitudes agrees with the results obtained in Sec. \ref{sec2}. The agreement is explicitly verified up to the four photon function.
Finally, in Sec. \ref{sec4} we discuss the main results.

\section{Thermal amplitudes from the Boltzmann transport equation}\label{sec2}
Let us consider the semiclassical description of a QED plasma in terms of a distribution 
\begin{equation}\label{eq1}
f(x,k,e) = \frac{1}{(2\pi)^3}2 \theta(k_0) \delta(k^2-m^2) F(x,k,e) ,
\end{equation}
where $x=(t,\vec x)$ and $k=(k_0,\vec k)$ are the four-vectors of position and momenta.  
The Dirac delta and theta functions impose the conditions  $k=(m/\sqrt{1-v^2},m \vec v/\sqrt{1-v^2})$ for
on-shell, positive energy, particles of charge $e$ and mass $m$. 
%The function $F(x,k)$ is to be determined.

Let us now assume that the collisions between the particles are neglectable. Then,
when an external electromagnetic field is present, the trajectory followed by the particles is determined by the equations of motion
\begin{equation}\label{eq2}
m \frac{dx^\mu}{d\tau} = k^\mu; \;\;\; m \frac{d k_\mu}{d\tau} = e F_{\mu\nu}k^\nu ,
\end{equation}
where $F_{\mu\nu} = \partial_\mu A_\nu - \partial_\nu A_\mu$ is the electromagnetic field tensor and $\tau$ is the proper time.
As a result, the distribution function satisfies the equation
\begin{equation}\label{eq3}
\frac{d}{d\tau} f(x,k,e) = \frac{\partial f}{\partial x^\mu}\frac{d x^\mu}{d\tau}+
                         \frac{\partial f}{\partial k_\mu}\frac{d k_\mu}{d\tau}  =  0.
\end{equation}
%where the vanishing righthand side results from the absence of collisions.
We remark that the right-hand side of the previous equation does not vanish when the effect of collisions is taken into account. 
Using Eqs. \eqref{eq2}, we obtain the Boltzmann transport equation 
\begin{equation}\label{eq4}
k \cdot \partial f(x,k,e) = -{\cal F} \cdot \partial_k f(x,k,e), %; \;\;\; \partial_k\equiv \frac{\partial}{\partial k_\mu}
\end{equation}
where $\partial_k$ is the partial derivative in relation to the four momentum $k_\alpha $ and we have defined
%the four-force 
%${\cal F}_\mu$ is given by
\begin{equation}\label{eq5}
{\cal F}_\alpha = e  F_{\alpha\beta}  k^\beta .
\end{equation} 
This covariant form of the Boltzmann transport equation has been previously obtained in more general cases of non-Abelian gauge fields and
also noncommutative gauge theories \cite{Heinz:1985yq,Elze:1986hq,Elze:1989un,kelly:1994dh,Brandt:2002qk,Brandt:2002rw,Brandt:2003je}.

Let us now consider the current which is produced when the plasma interacts with the external field.
This can be expressed in terms of the distribution functions for the negative or the positive charge carriers (e.g. positrons or electrons)  as
\begin{equation}\label{eq7}
{}^{\pm}\!\! J_\mu =  \pm e\, {\cal C} \int d^4 k \, k_\mu f(x,k,\pm e) .
\end{equation}
where ${\cal C}$ takes into account the degrees of freedom associated with the charged particles 
(${\cal C} = 2$ for an electron in 3+1 dimensions).
%which can also be written as 
The total current in a plasma would be given by
%In an neutral plasma the full current would be given by
\begin{equation}\label{eq6}
J_\mu = {}^{+}\!J_\mu + {}^{-}\!J_\mu .
\end{equation}
%are the partial contributions of positrons and electrons. 
%The special choice made in Eq. \eqref{eq6} ensures the plasma neutrality. 
Of course one could think of other possible scenarios such that the net charge of the plasma does not vanish. For this reason, it is convenient to consider 
${}^{\pm}\!J_\mu$ separately.

If we now perform the replacement $k\rightarrow -k$ in Eq. \eqref{eq4}, we obtain
\begin{equation}\label{eq8}
k \cdot \partial f(x,-k,e) = +{\cal F} \cdot \partial_k f(x,-k,e).
\end{equation}
On the other hand, if we make $e \rightarrow -e$ we get
\begin{equation}\label{eq9}
k \cdot \partial f(x,k,-e) = +{\cal F} \cdot \partial_k f(x,k,-e).
\end{equation}
This shows that we can  choose  $f(x,k,-e) = f(x,-k,e)$. 
Therefore we can, alternatively, express Eq. \eqref{eq7} as
%make the identification (without loss of generality)  
%$f(x,k,-e) = f(x,-k,e)$ 
\begin{equation}\label{eq10}
{}^{\pm}\!\! J_\mu = \pm e \,{\cal C} \int d^4 k \, k_\mu f(x,\pm k, e) ,
\end{equation}
so that the charge conjugation is equivalent to reversing the sign of the thermal particle momentum.

%
%It is worth noticing that under the transformation $k\rightarrow -k$, Eq. \eqref{eq4}  becomes
%\begin{equation}\label{eq8}
%k \cdot \partial f(x,-k,e) = +{\cal F} \cdot \partial_k f(x,-k,e).
%\end{equation}
%On the other hand, if we make $e \rightarrow -e$ we obtain
%\begin{equation}\label{eq9}
%k \cdot \partial f(x,k,-e) = +{\cal F} \cdot \partial_k f(x,k,-e).
%\end{equation}
%Therefore we can make the identification (without loss of generality)  $f(x,k,-e) = f(x,-k,e)$, so that
%\begin{equation}\label{eq10}
%{}^{\pm}\!\!J_\mu = \pm e \int d^4 k  f(x,\pm k, e)  k_\mu .
%\end{equation}

%In order to make contact with thermal field theory we 
Let us now consider the general relation that must exist between the current and the thermal effective action of electromagnetic fields interacting with a thermal plasma. This relation can be simply written as
\begin{eqnarray}\label{eq11}
{}^{\pm}\! \Gamma_{transp} &=& \int d^4 x  \; {}^{\pm}\!\! J_\mu(A)  \,A^\mu(x) , 
\end{eqnarray}
where we have made explicit the dependence of  ${}^{\pm}\!\!J_\mu$  
on the external field $A_\mu$, which follows from 
the substitution of the solution of Eq. \eqref{eq4} into Eq. \eqref{eq10}.

In the most trivial scenario, when the external field is switched off, and the system is in thermal equilibrium,
Eq. \eqref{eq4} is satisfied by
\begin{equation}\label{eq12}
f(x,k,e) = f^{(0)}(k) = \frac{1}{(2\pi)^3}2 \theta(k_0) \delta(k^2-m^2) F^{(0)}(k_0).
\end{equation}
The equilibrium distribution $F^{(0)}(k_0)$ depends on the statistics of the particles. In the case of electrons and positrons 
\begin{equation}\label{eq13}
F^{(0)}(k_0) = N_F(k_0) = \frac{1}{\exp{(k_0/T)}+1} 
\end{equation} 
which is the Fermi-Dirac distribution (for our present purposes $F^{(0)}(k_0)$ could be any $x$-independent distribution).
Then, the partial currents are given by
\begin{eqnarray}\label{eq14}
{}^{\pm}\!\!J_\mu ={}^{\pm}\!\! J^{(0)}_\mu &=& \pm
\frac{e \,{\cal C}}{(2\pi)^3} \int \frac{d^3 k}{\sqrt{|\vec k|^2+m^2}} 
\nonumber \\ &{} & N_F\left(\sqrt{|\vec k|^2+m^2}\right) k_\mu
\end{eqnarray}
and the corresponding effective actions are 
\begin{eqnarray}\label{eq15}
{}^{\pm}\! \Gamma_{transp}^{(1)} &=& \pm \frac{e\,{\cal C}}{(2\pi)^3} \int d^4 x \int \frac{d^3 k}{\sqrt{|\vec k|^2+m^2}} 
\nonumber \\ &{} & N_F\left(\sqrt{|\vec k|^2+m^2}\right) k_\mu A^{\mu}(x) .
\end{eqnarray}
%In the high temperature limit, when $m\ll T$, we obtain
%\begin{eqnarray}\label{eq16}
%{}^{\pm}\! \Gamma_{transp}^{(1)} &=& \pm \frac{e\,{\cal C}}{(2\pi)^3} \int d^4 x \int \frac{d^3 k}{|\vec k|} 
%\nonumber \\ &{} & N_F\left(|\vec k|\right) k_\mu A^{\mu}(x) .
%\end{eqnarray}
Introducing the four-vector $K=(\sqrt{1 + m^2/|\vec k|^2},\vec k/|\vec k|)$, we can write
\begin{eqnarray}\label{eq17}
  {}^{\pm}\! \Gamma_{transp}^{(1)} &=&  \pm \frac{{\cal C} e T^3  I^{(1)}(\bar m) }{(2\pi)^3}  
\nonumber \\ &{} &  \int d^4x \int d\Omega \, K_\mu A^\mu(x)  ,
\end{eqnarray}
where $\int d\Omega$ denotes the integration over the directions of  $\hat K = \vec k/|\vec k|$ and we have introduced
the integration operator 
\begin{equation}
I^{(n)}(\bar m) \equiv \int_0^{\infty} du  \frac{u^{4-n}}{\sqrt{u^2+ {\bar m}^2}}  \frac{1}{{\rm e}^{\sqrt{u^2+ {\bar m}^2}} +1}
%\dots
;\;\;\; \bar m \equiv \frac{m}{T}  ,
\end{equation}
which operates on functions of $K_0=\sqrt{1+\bar m^2/u^2}$.
(together with the powers of the temperature, $I^{(n)}(\bar m)$ give the full 
temperature dependence for the contribution of order $n$).
Since $\int d\Omega \, K_i = 0$, we finally obtain
\begin{eqnarray}\label{eq18}
  {}^{\pm}\! \Gamma_{transp}^{(1)} &=&  \pm \frac{e \,{\cal C}T^3}{2\pi^2} 
\int_0^{\infty} du  \frac{u^{2}}{{\rm e}^{\sqrt{u^2+ {\bar m}^2}} +1}
%I^{(1)} \sqrt{1+\frac{\bar m^2}{u^2}}
% \int_0^{\infty} du \frac{u^2}{{\rm e}^u+1}  
\int d^4x  A^0(x)  . \nonumber \\ &&
\end{eqnarray}  
%where we have used Eq. \eqref{b2a}.
%\begin{equation}\label{eq19}
%\int_0^{\infty}  \frac{u^\alpha du}{{\rm e}^u+1} = \left(1-\frac{1}{2^\alpha}\right) \Gamma(\alpha+1)\zeta(\alpha+1).   
%\end{equation}
Of course, in the case of a neutral plasma, the sum of the two partial contributions vanishes 
\begin{equation}\label{eq20}
  \Gamma_{transp}^{(1)} =  {}^{+}\!\Gamma_{transp}^{(1)}  +  {}^{-}\!\Gamma_{transp}^{(1)}  = 0. 
\end{equation}
As we will see this simple property will remain true for all odd powers of the external field (see Eq. \eqref{eq26} ).

In the presence of a general external field, the solution of the Boltzmann transport equation can be quite involved. 
Here we consider external field configurations which can be dealt with using 
a perturbative approach. In these cases a formal solution of the Eq. \eqref{eq4} can be written as
\begin{equation}\label{pertf}
f^{(n)}(x,\pm k, e)  = \left(\mp \frac{1}{k\cdot \partial }{\cal F} \cdot \partial_k \right)^n f^{(0)}(k),
\end{equation}
%The perturbative solutions which exhibit an non-trivial dependence on the external fields can be written as
so that
%\begin{equation}
%f(x,k,e) = f^{(0)}(k) + f^{(1)}(x,k,e) + f^{(2)}(x,k,e) + \dots ,
%\end{equation}
each $f^{(n)}$ is of order $n$ in powers of the external field and the complete phase space distribution
can be expressed as $f(x,k,e) = f^{(0)}(k) + f^{(1)}(x,k,e) + \dots$. 
It is convenient to write Eq. \eqref{pertf} in a more explicitly iterative fashion as
\begin{equation}\label{eq21}
f^{(n)}(x,\pm k,e) = \mp e\;  \frac{1}{k\cdot\partial} F_{\alpha\beta}  k^\beta  \frac{\partial}{\partial k_\alpha} f^{(n-1)}(x,\pm k, e),
\end{equation}
where $n=1,\;2,\; \dots$.  
From these solutions, and using Eqs. \eqref{eq10} and \eqref{eq11},  
we can obtain the corresponding partial effective actions, 
defined order by order in terms of the phase space distributions, as follows 
%\begin{equation}\label{eq23}
%{}^{\pm}\!\!J^{(n-1)}_\mu = \pm e \int d^4 k k_\mu f^{(n-1)}(x,k,\pm e); \;\; n=1,\, 2,\, \dots .
%\end{equation}
%And the effective actions
\begin{equation}\label{eq22}
{}^{\pm}\! \Gamma_{transp}^{(n)}  = \pm  {e\,{\cal C}}\int d^4 x \int d^4 k \, A^\mu \, k_\mu \,f^{(n-1)}(x,\pm k, e)
\end{equation}
%(this is the order $n$ term in the Taylor series expansion of the effective action, so that the
%functional derivaties of ${}^{\pm}\! \Gamma_{transp}$ are identified with 1PI amplitudes).
In principle, the effective action is completely determined by this direct procedure starting from the knowledge of
$f^{(0)}$, which can be chosen, for instance, as given by Eqs. \eqref{eq12} and \eqref{eq13}.  

The nontrivial effects of the external field interacting with the plasma will be manifest when we consider the second or
higher order contributions in Eq. \eqref{eq22} ($n\ge 2$). 
To derive the explicit dependence on the external field, let us look at the details of the integrand in Eq. \eqref{eq22}.
Substituting Eq. \eqref{eq21} into  Eq. \eqref{eq22} we obtain 
%one can see that this integrand involves the
%structure {\bf\color{red} [indices de F estavam trocados]}
\begin{eqnarray}\label{eq23}
{}^{\pm}\! \Gamma_{transp}^{(n)}  &=&  -{e^2 \,{\cal C}}\int d^4 x \int d^4 k \,  A^\mu k_\mu \frac{1}{k\cdot\partial} k^\beta F_{\alpha\beta} 
\nonumber \\ &&
\frac{\partial}{\partial k_\alpha} f^{(n-2)}(x,\pm k,e).
\end{eqnarray}
Performing an integration by parts in the momentum integral, yields
\begin{eqnarray}\label{eq24}
{}^{\pm}\! \Gamma_{transp}^{(n)}  &=&  {e^2 \,{\cal C}}\int d^4 x \int d^4 k \,  
A^\mu \left( \delta^\alpha_\mu \frac{1}{k\cdot\partial} - \frac{k_\mu \partial^\alpha}{(k\cdot\partial)^2} \right) 
\nonumber \\
&&k^\beta F_{\alpha\beta} f^{(n-2)} 
\nonumber \\
&=&  {e^2 \,{\cal C}}\int d^4 x \int d^4 k \,  
A^\mu \left( \delta^\alpha_\mu k_\lambda\partial^\lambda - k_\mu \partial^\alpha\right) 
\nonumber \\
&&\frac{1}{(k\cdot\partial)^2}   k^\beta F_{\alpha\beta} f^{(n-2)} ,
\end{eqnarray}
where the derivative of $\partial/\partial k_\alpha$ acting on $k^\beta$ produces a vanishing contribution
as a consequence of the antisymmetry of the electromagnetic tensor field.
Performing an integration by parts in the configuration space
integral and using again the antisymmetry of the electromagnetic tensor field %in order to reverse a sign,
we obtain % {\bf\color{red} [indices do segundo F estavam trocados; o primeiro esta correto]}
%\begin{equation}\label{eq25}
%k_\lambda F^{\lambda\alpha} \frac{1}{(k\cdot\partial)^2} k^\beta F_{\alpha\beta} f^{(n-2)} .
%\end{equation}
%Therefore, Eq. \eqref{eq22} can be written as {\bf\color{red} [sinal menos incluido devido a troca dos indices de um dos Fs]}
\begin{eqnarray}\label{eq26}
{}^{\pm}\! \Gamma_{transp}^{(n)} &=&   {e^2 \,{\cal C}}\int d^4 x \int d^4 k 
F^{{\mu_2}{\nu_2}} \frac{1}{(k\cdot\partial)^2} F^{{\mu_1}{\nu_1}}  
\nonumber \\ &{}& { k_{\mu_1}  k_{\mu_2} \eta_{\nu_1\nu_2}} f^{(n-2)}(x, \pm k,  e) 
\end{eqnarray}
(from this result one can easily verify that, for odd $n$,  ${}^{+}\! \Gamma_{transp}^{(n)}  + {}^{-}\! \Gamma_{transp}^{(n)}  = 0 $).

This form of the effective action encodes a number of interesting features. 
First, since Eq. \eqref{eq26} is expressed directly in terms of $F_{\mu\nu}$ (notice that $f^{(n-2)}$ is implicitly dependent on $F^{\mu\nu}$ as a solution of \eqref{eq21}), gauge invariance is explicitly manifested at any given order.
%As a result, the amplitudes will satisfy simple Ward identities.
Second, it is straightforward to obtain any higher order contributions by successive substitutions of Eq. \eqref{eq21}  into Eq. \eqref{eq26}. 
These properties can be concisely summarized by expressing the external fields 
in terms of Fourier components so that the effective actions can be written as
\begin{eqnarray}\label{eq26a}
{}^{\pm}\! \Gamma_{transp}^{(n)} &=&  \frac{1}{n!}  \int \frac{d^4 p_1}{(2\pi)^4}  \frac{d^4 p_2}{(2\pi)^4} \dots \frac{d^4 p_n}{(2\pi)^4}   
\nonumber \\ &   {} & \tilde A^{\mu_1}(p_1)   \dots\tilde  A^{\mu_n}(p_n)   
\nonumber \\ &   {} &  {}^{\pm}  \Pi^{transp}_{\mu_1\mu_2 \dots \mu_n}(p_1,p_2,\dots,p_n)  
\nonumber \\ &   {} &  (2\pi)^4 \delta(p_1+\dots+ p_n)  , 
\end{eqnarray}
where ${}^{\pm}  \Pi^{transp}_{\mu_1\mu_2 \dots \mu_n}(p_1,p_2,\dots,p_n)$ are the thermal amplitudes associated with the Boltzmann transport equation
formalism. Then, from the gauge invariance of ${}^{\pm}\! \Gamma_{transp}^{(n)}$  
under $\tilde A^{\mu_i} \rightarrow \tilde A^{\mu_i} + \tilde\Lambda p_i^{\mu_i}$, we conclude that the thermal amplitudes must satisfy
the Ward identities
\begin{equation}
p_i^{\mu_i} \Pi^{transp}_{\mu_1\mu_2 \dots \mu_n}(p_1,p_2,\dots,p_n)  = 0.
\end{equation}  
Notice that in the case of non-Abelian gauge theories %\cite{natransp},  
the corresponding Ward identities would relate 
$\Pi^{transp}_{\mu_1\mu_2 \dots \mu_n}(p_1,p_2,\dots,p_n) $ with $\Pi^{transp}_{\mu_1\mu_2 \dots \mu_{n-1}}(p_1,p_2,\dots,p_{n-1})$ so that
all the amplitudes would have the same leading high temperature behavior.
On the other hand, in the Abelian theory, each contribution ${}^{\pm}\! \Gamma_{transp}^{(n)} $ to the effective action is individually gauge independent and the corresponding amplitudes are not all proportional to the same power of the temperature. This can be simply understood by power counting the momentum dependence of successive terms obtained from Eq. \eqref{eq21}. Furthermore, from the general form of Eq. \eqref{eq26} one can easily see that all the amplitudes will be functions of degree zero in the external momenta $p$.

As an example, let us now compute the explicit form of the two-photon amplitude (thermal self-energy). Making $n=2$ in Eq. \eqref{eq26} one readily obtains
 %{\bf\color{red} [sinal menos incluido devido a troca dos indices de um dos Fs]}
\begin{eqnarray}\label{eq27}
{}^{\pm}\! \Gamma_{transp}^{(2)} &=&  {e^2 \,{\cal C}} \int d^4 x \int d^4 k 
F^{{\mu_2}{\nu_2}} \frac{1}{(k\cdot\partial)^2} F^{{\mu_1}{\nu_1}}  
\nonumber \\ &{}& { k_{\mu_1}  k_{\mu_2} \eta_{\nu_1\nu_2}} f^{(0)} .
\end{eqnarray}
Using Eqs. \eqref{eq12} and \eqref{eq13} % and neglecting the mass $m$, we obtain 
 %{\bf\color{red} [sinal menos incluido devido a troca dos indices de um dos Fs]}
\begin{eqnarray}\label{eq28}
{}^{\pm}\! \Gamma_{transp}^{(2)} &=&  \frac{e^2 {\cal C} T^2 I^{(2)}(\bar m) }{(2\pi)^3}  \int d^4 x  %\int_0^\infty  \frac{u du}{{\rm e}^u+1}
\nonumber \\ &{}&  \int d\Omega
F^{{\mu_2}{\nu_2}} \frac{ K_{\mu_1}  K_{\mu_2} \eta_{\nu_1\nu_2} }{(K\cdot\partial)^2} F^{{\mu_1}{\nu_1}}  .
\end{eqnarray}
Expressing the external fields in terms of Fourier components we obtain
 %{\bf\color{red} [sinal menos incluido devido a troca dos indices de um dos Fs]}
\begin{eqnarray}\label{eq29}
{}^{\pm}\! \Gamma_{transp}^{(2)} &=&  \frac{ e^2 {\cal C} T^2 I^{(2)}(\bar m)}{(2\pi)^3} \int \frac{d^4 p}{(2\pi)^4} %\int_0^\infty  \frac{u du  }{{\rm e}^u+1}
\nonumber \\ &{}&  \int d\Omega
 \frac{\tilde F^{{\mu_2}{\nu_2}}  K_{\mu_1}  K_{\mu_2} \eta_{\nu_1\nu_2}  \tilde F^{{\mu_1}{\nu_1}}  }{(K\cdot p)^2},
\end{eqnarray}
where 
\begin{equation}\label{eq30}
\tilde F^{\mu_i\nu_i} \equiv %p^{\mu_i} \tilde A^{\nu_i}(p) - p^{\nu_i} \tilde A^{\mu_i}(p) = 
\left( p^{\mu_i} \eta^{\nu_i\mu} - p^{\nu_i} \eta^{\mu_i\mu}  \right) A_\mu .
\end{equation}
Inserting \eqref{eq30} into \eqref{eq29} and comparing the resulting expression with
\eqref{eq26a}, we obtain the following result for the two-photon amplitude
% {\bf [sinal corrigido devido a troca dos indices de um dos Fs]}
\begin{eqnarray} \label{eq33}
 {}^{\pm}\!\Pi^{transp}_{\mu\nu}  & = &  -\frac{2 e^2 {\cal C} T^2I^{(2)}(\bar m) }{(2\pi)^3} %\int_0^\infty  \frac{u du}{{\rm e}^u+1}
%\frac{\pi^2}{6}  
\int d\Omega
\nonumber \\ &{} & \!\!\!\!\!\! \!\!\!\!\!\!\!\!\! \!\!\!
\left(
\eta_{\mu\nu} - \frac{K_\mu p_\nu + K_\nu p_\mu}{K\cdot p} + \frac{p^2 K_\mu K_\nu}{(K\cdot p)^2} 
\right) .
%\nonumber \\
\end{eqnarray}
%where we have used Eq. \eqref{eq19} with $\alpha=1$.
This particular example of the two-photon function illustrates the general structure of 
all $n$-photon functions, which, as we will see can be expressed 
in terms of an angular integral over the directions of \hbox{$\hat K \equiv {\vec k}/{|\vec k|}$}, 
the integrand being a rank $n$ tensor which depends on $K_\mu$ as well as the external momenta.
The same structure has been previously obtained for the leading high temperature limit of non-Abelian as well as 
noncommutative gauge theories \cite{Brandt:2002rw}.

Iterating  Eq. \eqref{eq21} one more time, Eq. \eqref{eq26} yields % {\bf\color{red}[Verifiquem os fatores!]} 
\begin{eqnarray}\label{eq34}
{}^{\pm}\! \Gamma_{transp}^{(n)} &=&  \mp {e^3} {\cal C} \int d^4 x \int d^4 k   
F^{{\mu_2}{\nu_2}} \frac{k_{\mu_1}  k_{\mu_2} \eta_{\nu_1\nu_2} }{(k\cdot\partial)^2} F^{{\mu_1}{\nu_1}}   
\nonumber \\
&{}& \left(k_\beta \frac{1}{k\cdot\partial} F^{\alpha \beta} \frac{\partial }{\partial k^{\alpha} } \right)  f^{(n-3)} (x,\pm k, e) .
\end{eqnarray}
Performing an integration by parts in the momentum integral, the result can be cast 
in the form
\begin{widetext}
\begin{eqnarray}\label{eq36}
{}^{\pm}\! \Gamma_{transp}^{(n)} =  \pm {e^3} {\cal C} \int d^4 x \int d^4 k   F^{{\mu_1}{\nu_1}} \frac{\partial^{\lambda_1}}{(k\cdot\partial)^3} 
F^{{\mu_2}{\nu_2}}  \frac{\partial^{\lambda_2} \partial^{\lambda_3} }{(k\cdot\partial)^3} F^{{\mu_3}{\nu_3}} \eta_{\nu_1\nu_2}
T^3_{\mu_1\mu_2\mu_3\nu_3\lambda_1\lambda_2\lambda_3}     f^{(n-3)} (x,\pm k, e) ,
\end{eqnarray}  
where % {\bf[\color{red} VERIFIQUEM SE ESTE TENSOR ESTA CORRETO]}
\begin{eqnarray}\label{eq37}
T^3_{\mu_1\mu_2\mu_3\nu_3\lambda_1\lambda_2\lambda_3}     &=& 2 k_{\mu_1} k_{\mu_2} k_{\mu_3} k_{\lambda_2} k_{\lambda_3} \eta_{\lambda_1\nu_3}
%\nonumber \\ & - & 
-\left[  k_{\mu_1} k_{\mu_3} k_{\lambda_1} k_{\lambda_2} k_{\lambda_3} \eta_{\mu_2\nu_3} + \mu_1 \leftrightarrow \mu_2 \right]
%\nonumber \\ & + & 
+k_{\mu_1} k_{\mu_2} k_{\mu_3} k_{\lambda_1} k_{\lambda_2} \eta_{\lambda_3\nu_3}
\end{eqnarray}
\end{widetext}
It is then straightforward to obtain the three-point amplitudes. 
Similarly to the previous cases, we now consider $n=3$ 
in the Eq. \eqref{eq36} and express the external fields in terms of Fourier components. 
Comparing the resulting expression with Eq. \eqref{eq26a} we obtain the result
\begin{eqnarray}\label{eq38}
  {}^{\pm}\! \Pi^{transp}_{\mu_1\mu_2\mu_3} &=&  \pm \frac{6 e^3 {\cal C} T I^{(3)}(\bar m)}{(2\pi)^3}  %\int_0^{\infty}  \frac{du}{{\rm e}^u+1}  
%\nonumber \\ &{} & 
\int d\Omega \; {\cal A}_{\mu_1\mu_2\mu_3} .
\end{eqnarray}
where the explicit result for ${\cal A}_{\mu_1\mu_2\mu_3}$ is given in Appendix \ref{apa}.
%The numerical factor $\log(2)$ is the limit of Eq. \eqref{eq19} when $\alpha\rightarrow 0$. 
%In fact, the three point function
%is the threshold case, since for higher point functions ($n\ge 4$) Eq. \eqref{eq19} will no longer 
%be applicable and it will be necessary to introduce an infrared cutoff for the momentum of the thermal particles. 

Using this iterative approach, we have also computed the explicit result for the four-photon amplitude. Since the calculation is straightforward, but rather
involved, we have made use of computer algebra. The result has been compared with the one obtained using the high temperature limit of thermal field theory, as we
describe in the next section.

It is clear that, in general, the iterative procedure will produce a result for the $n$-photon function which is proportional to
\begin{equation}\label{eq38a}
\frac{(\pm e)^n {\cal C}  T^{4-n} I^{(n)}(\bar m) }{(2\pi)^3}   \int d\Omega \; {\cal A}_{\mu_1\mu_2\dots\mu_n} 
\end{equation}
so that each thermal amplitude has a simple temperature dependence given by the factors $T^{4-n} I^{(n)}(\bar m)$.

%{\bf \color{red} secao ainda nao terminada ... confiram tudo até aqui (inclusive o texto!)}

\section{Hard thermal loops in QED}\label{sec3}
In this section we consider the QED $n$-photon amplitudes in the usual thermal field theory setting \cite{kapusta:book89,lebellac:book96,das:book97}.
Furthermore, we chose the imaginary time formalism, which, as it will be seen, is the
most direct approach to obtain the thermal amplitudes in the present context.
%In general the thermal amplitudes can be computed 
%Examples of Feynman diagrams which contribute to these amplitudes are shown in figure \ref{fig1} in the context of QED. 
Also, to fully describe a plasma under the influence of an external electromagnetic fields, the external energies in 
these diagrams have to be analytically continued to continuous values, after the Matsubara sums have been performed.
This prescription extends the imaginary time formalism to a nonequilibrium regime under the influence of external fields.

\begin{figure}[t!]
\qquad\includegraphics[scale=0.2]{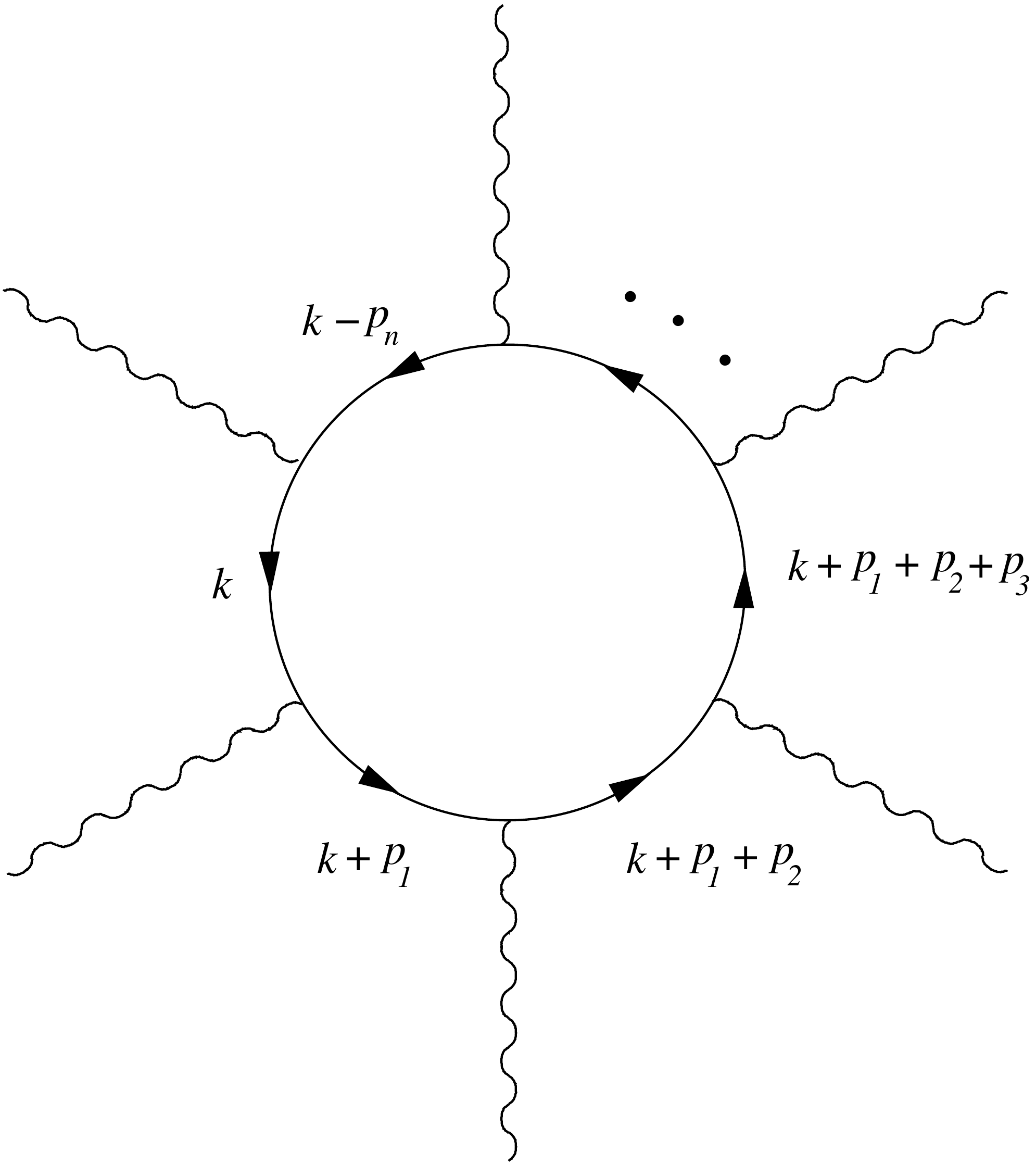} 
% \\       
%\includegraphics[scale=0.3]{cubic} 
%\qquad\includegraphics[scale=0.3]{quartic}
\caption{One-loop amplitude which contribute to the effective action.
The solid and wavy lines represent fermions and photons respectively.
The external momenta are denoted by $p_i$, $i=1,\dots n$ and $k$ denotes the loop momenta.
The symmetrized amplitude which contributes to the effective action is obtained by adding all the permutations of $(n-1)$ 
external photon lines, and dividing the result by $(n-1)!$ .}
\label{fig1}
\end{figure}

Let us consider the one-loop diagram, shown in Fig. \ref{fig1}. Once we compute this basic diagram, the $n$-point one-particle-irreducible (1PI) amplitudes 
can be obtained as the sum of certain permutations of the pairs $(p_i,\;\mu_i)$, so that the bosonic symmetry is satisfied.
%For instance, the  four point 1PI amplitude is the sum of three permutations of the basic amplitude,  so that the bosonic symmetry is satisfied.
In the imaginary time formalism the contribution of this basic diagram can be written as %(for simplicity, we are omitting the Lorentz indices).
\begin{equation}\label{eq31}
{G}_{\mu_1\mu_2\dots\mu_n} = -
T \int\,\frac{{\rm d}^3  k}{(2\pi)^3}  \sum_{k_0=i\,\omega_n} f_{\mu_1\mu_2\dots\mu_n}(k_0,\vec k) %  \, {\cal S}_n(\vec k;p_1,\cdots,p_n),
\end{equation}
where
\begin{widetext}
\begin{equation}\label{integ1}
f_{\mu_1\mu_2\dots\mu_n} (k_0,\vec k) = % {\cal S}_n\equiv T\, \sum_{k_0=i\,\omega_n}
\frac{1}{k_0^2-|\vec k|^2 - m^2 }\frac{1}{(k_0+{p_1}_0)^2-(\vec k+\vec p_1)^2-m^2}
\cdots 
\frac{t_{\mu_1\mu_2\dots\mu_n}(k;p_1,\cdots,p_n)}{(k_0-{p_n}_0)^2-(\vec k-\vec p_n)^2-m^2}.
\end{equation}
\end{widetext}
and we have taken into account the minus sign associated with the fermion loop.
The factor $t_{\mu_1\mu_2\dots\mu_n}(k;p_1,\cdots,p_n)$ is a shorthand notation for all the contributions to the numerator which 
arises, for instance, from the traces of Dirac matrices and coupling constant factors.
The quantities $\omega_n=(2n+1) \pi T$ are the fermionic Matsubara frequencies. % ($\omega_n = 2 n \pi T$ for bosons).
Using the identity \cite{kapusta:book89,lebellac:book96,das:book97}
\begin{eqnarray}\label{contour1}
T\displaystyle{\sum_{n=-\infty}^{\infty}}
f(k_{0}=i\,\omega_n)&=&
\displaystyle{1\over 2\pi i}\displaystyle{\oint_{C}}d k_{0}f(k_{0})
\nonumber \\ &&
{1\over 2}\left[\tanh\left({1\over 2}\beta k_{0}\right)\right],
\end{eqnarray}
it is straightforward to show that the temperature dependent part of \eqref{eq31} can be written as
% sinal de soma = 1/2 - N_f
\begin{eqnarray}\label{TermVac}
G^{therm}_{\mu_1\mu_2\dots\mu_n}  &=& 
\int\,\frac{{\rm d}^3  k}{(2\pi)^3}
\int_{-i\infty+\delta}^{i\infty+\delta}
\frac{{\rm d} k_0}{2\pi i} N_{F}(k_0)
\nonumber \\ &{} &
\left[f_{\mu_1\mu_2\dots\mu_n} (k_0,\vec k)+f_{\mu_1\mu_2\dots\mu_n} (-k_0,\vec k)\right].
\nonumber \\ & &
\end{eqnarray}
Making the change of variables $\vec k \rightarrow -\vec k$ in the second term inside the bracket and using
the shorthand notation $f_{\mu_1\mu_2\dots\mu_n}(k_0,\vec k) = f_{\mu_1\mu_2\dots\mu_n}(k)$, we can write the thermal contribution of the $n$-photon diagram in terms of two components, as
\begin{equation}\label{components}
G^{therm} _{\mu_1\mu_2\dots\mu_n}  =  {}^{+}\! G^{therm} _{\mu_1\mu_2\dots\mu_n}   +  {}^{-}\! G^{therm} _{\mu_1\mu_2\dots\mu_n}   ,
\end{equation}
where
\begin{eqnarray}\label{TermVac1}
{}^{\pm}\! G^{therm} _{\mu_1\mu_2\dots\mu_n} &=&    
\int\,\frac{{\rm d}^3  k}{(2\pi)^3}
\int_{-i\infty+\delta}^{i\infty+\delta}
\frac{{\rm d} k_0}{2\pi i} N_{F}(k_0)
\nonumber \\ &{} &
%\left[
f_{\mu_1\mu_2\dots\mu_n} (\pm k).
%\right].
\end{eqnarray}
This separation in two components is the first necessary step in order to compare the field theory formalism with the results obtained
in the previous section.

\begin{figure}[t!]
\qquad\includegraphics[scale=0.4]{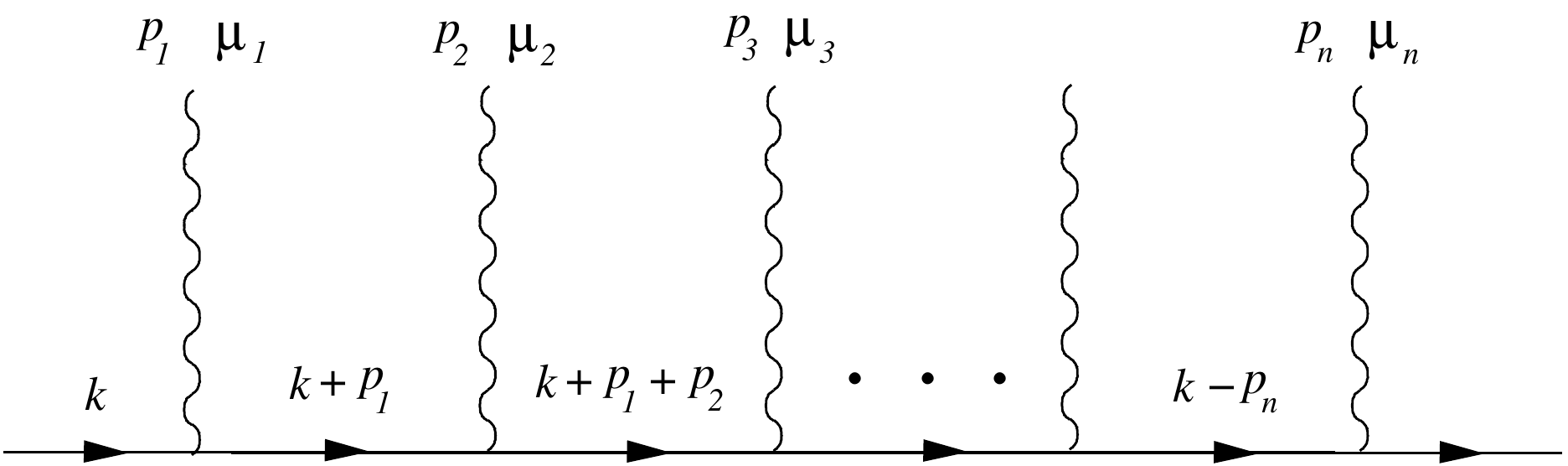} 
% \\       
%\includegraphics[scale=0.3]{cubic} 
%\qquad\includegraphics[scale=0.3]{quartic}
\caption{Diagram representing the forward scattering amplitude ${A}^n_{\mu_1\mu_2\dots\mu_n} (k,p_1,\cdots,p_n)$.
%The forward scattering amplitude of thermal particles associated with the amplitude in Fig.\ref{fig1}. 
A trace over the product of gamma matrices is to be understood.}
%Cyclic permutations of the external lines are implied.}
\label{fig2}
\end{figure}

Closing the contour of the $k_0$ integration in Eq. \eqref{TermVac1} in the 
right-hand side of the complex plane and taking into account the poles from the denominator of Eq. \eqref{integ1} we obtain
% sinal do sentido horário
\begin{eqnarray}\label{forward1}
{}^{\pm}\! G^{therm} _{\mu_1\mu_2\dots\mu_n} &=&   -\int\,\frac{{\rm d}^3 k}{(2\pi)^3} 
\frac{N_{F}\left(\sqrt{(\vec k)^2+m^2} \right) }{2\sqrt{(\vec k)^2+m^2}}
\nonumber \\ &{}&
%\left[ 
\sum_{{cycl.}} {A}^n_{\mu_1\mu_2\dots\mu_n}(\pm k,p_1,\cdots,p_n), %%%%% + k\leftrightarrow -k\right],
\end{eqnarray}
where the quantities ${A}^n_{\mu_1\mu_2\dots\mu_n}$ denote on-shell \hbox{($k_0=\sqrt{(\vec k)^2+m^2}$)}
forward scattering amplitudes, as depicted in Fig. \ref{fig2}, and $\sum_{{cycl.}}$
denotes the sum of all cyclic permutations of the pairs $(p_i,\;\mu_i)$, as we will explain soon. 

At this point, we analytically continue the result to the Minkowski space-time, so that the amplitudes
${A}^n_{\mu_1\mu_2\dots\mu_n}(k;p_1,p_2,\cdots,p_n)$ can be computed using the zero temperature QED Feynman rules. 
In this way, we obtain from Fig. \ref{fig2} the following result (an extra minus sign arises from the analytic continuation) 
%{\bf\color{red} [NOTE O SINAL MENOS DO LOOP FERMIONICO; 
%Agora estamos usando regras de Feynman que não introduzem qualquer sinal extra}
%TAMBEM UM FATOR $i$ para cada vertice e $i$ para cada propagador... logo $(-1)^n$]}
\begin{subequations}\label{amp1}
\begin{eqnarray}
A^1_{\mu_1} &=& -e \, \tr (\gamma_{\alpha_1} \gamma_{\mu_1}) k^{\alpha_1} = -4e\, k_{\mu_1} \\ %\nonumber \\  
{A}^n_{\mu_1\dots\mu_n} %(k;p_1,p_2,\cdots,p_n) 
&=& -  e^n \,\left(\tr\, \gamma_{\alpha_1} \gamma_{\mu_1}  \dots \gamma_{\alpha_n} \gamma_{\mu_n}\right)
\nonumber \\ & {} & \!\!\!\!\!\!\!\!\!\!\!\!
\frac{k^{\alpha_1}  \dots (k+s_{n-1})^{\alpha_n} }
{(p_1^2+2k\cdot p_1)  \dots (s_{n-1}^2+2k\cdot s_{n-1})}, 
\end{eqnarray}
\end{subequations}
where
\begin{equation}
s_{n-1} \equiv p_1+p_2+\dots + p_{n-1}; \;\;\; n\ge 2 .
\end{equation}
We remark that, as a consequence of the on-shell condition, all the even powers of $m$ cancel. Also the terms proportional to
odd powers of $m$ vanish  simply because they are proportional to the trace of an odd number of gamma matrices, which is zero when
the space-time dimension is even (in an odd-dimensional space-time there would be induced Chern-Simons terms proportional to the odd
powers of $m$).
This way of expressing the $n$-photon amplitudes in terms of traces of $2n$ gamma matrices is convenient in order to carry a computer
algebra calculation.

%Of course it is implicit that in the special case when $n=1$ there is no denominator. 
%so that the amplitude is simply given by ${A}^1_{\mu}(k) = e\,{\rm tr} \slashed{k} \gamma_\mu = 4 e k_\mu$.
  %one has been used to perform the integration by residues

The particular set of forward scattering amplitudes given by Eq. \eqref{amp1} includes the 
ones which arise from the residue of the pole of $1/k^2$ in Eq. \eqref{integ1}.
It is straightforward to show that the corresponding contributions associated with each pole of Eq. \eqref{integ1} 
produce a result which is given by a cyclic permutation 
of the pairs $(p_i,\;\mu_i)$. In this way, the final result can  be expressed as
the sum $\sum_{{cycl.}}$ in Eq. \eqref{forward1}.
This representation of one-loop 1PI diagrams in terms of forward scattering amplitudes of on-shell thermal particles
has been derived in various other physical systems such as non-Abelian gauge theories and 
gravity \cite{Frenkel:1992ts,Brandt:1997se,Brandt:2000ht,Brandt:2006aj}. As a result, in all these cases it is natural 
to think in terms of a physical scenario where on-shell thermal particles of momentum $k$ are scattered
by the external photons of momenta $p_i$, $i=1,\dots, n$, as advanced by Barton \cite{Barton:1990fk}.
This is clearly in tune with the Boltzmann transport equation approach of the previous section.

The full contribution to the 1PI thermal Green's functions can now be obtained adding all the $(n-1)!$ permutations of the basic result in 
Eq. \eqref{forward1}.  %and dividing the result by $(n-1)!$.  
Then, taking into account that
$
\sum_{{perm.} \atop  (n-1)} \sum_{{cycl.}}     =  \sum_{{perm.} \atop n} 
$,
we can express the 1PI amplitudes as
\begin{eqnarray}\label{1pi1} 
{}^{\pm}\! \Pi^{therm} _{\mu_1\mu_2\dots\mu_n} &=&   
-\int\,\frac{{\rm d}^3 k}{(2\pi)^3}  \frac{N_{F}\left(\sqrt{(\vec k)^2+m^2} \right) }{2\sqrt{(\vec k)^2+m^2}}
\nonumber \\ &{} & 
\sum_{{perm.} \atop n}  {A}^n_{\mu_1\mu_2\dots\mu_n}(\pm k,p_1,\cdots,p_n) ,
\nonumber \\
\end{eqnarray}
and the corresponding effective actions are
\begin{eqnarray}\label{eat}
{}^{\pm}\! \Gamma_{therm}^{(n)} &=&  \frac{1}{(n-1)!}  \int \frac{d^4 p_1}{(2\pi)^4}  \frac{d^4 p_2}{(2\pi)^4} \dots \frac{d^4 p_n}{(2\pi)^4}   
\nonumber \\ &   {} &  \tilde A^{\mu_1}(p_1) \tilde A^{\mu_2}(p_2)  \dots\tilde  A^{\mu_n}(p_n)  
\nonumber \\ &   {} &  {}^{\pm}  \Pi^{therm}_{\mu_1\mu_2 \dots \mu_n}(p_1,p_2,\dots,p_n)  
\nonumber \\ &   {} &  (2\pi)^4 \delta(p_1+\dots +p_n)  .
\end{eqnarray}
(We remark that, due to the symmetry of the basic diagram in Fig. \ref{fig1}, the sum of the $(n-1)!$ 
permutations of external legs produces  completely bosonic symmetric amplitudes $ {}^{\pm}  \Pi^{therm}_{\mu_1\dots \mu_n}(p_1,\dots,p_n)$.)

In the previous section we have seen that all the effective actions \eqref{eq26a} have a simple temperature dependence of the form $T^{4-n} I^{(n)}(\bar m)$, 
as a result of Eq. \eqref{eq38a}. On the other hand, the thermal field theory effective actions \eqref{eat} have a more complicated temperature dependence. 
From a physical point of view, we expect that the Boltzmann transport equation approach is restricted to the high temperature limit, when the system approaches
a classical limit. This indicates that in the thermal field theory description, we have to consider the limit  
when $T$ is high compared with the energy-momentum scale of the external field. %xyxyxy
This amounts to consider the so-called hard thermal loop approximation, which in turn implies that the integration in \eqref{1pi1} is dominated by
the region where \hbox{$k_\mu \gg {p_i}_\mu$}.  Denoting the HTL limit of  ${}^{\pm}  \Pi^{therm}_{\mu_1 \dots \mu_n}(p_1,\dots,p_n)$
by ${}^{\pm}  \Pi^{htl}_{\mu_1 \dots \mu_n}(p_1,\dots,p_n)$, the equivalence of the two effective actions would then imply that
 %\begin{widetext}
\begin{equation}\label{eqv1}
%\frac{1}{(n-1)!}  
 {}^{\pm}  \Pi^{htl}_{\mu_1 \dots \mu_n}(p_1,\dots,p_n)  = 
% \frac{1}{n!} 
\frac 1 n  {}^{\pm}  \Pi^{transp}_{\mu_1\dots \mu_n}(p_1,\dots,p_n)  .
\end{equation}
%\end{widetext}
In QED this is a well-known result for the leading $T^2$ contribution to the two-photon amplitude.  
(In the case of non-Abelian gauge theories it can be proved that all the HTL $n$-gluon functions, which  have the same $T^2$ behavior,
can be generated using the Boltzmann transport equation approach \cite{kelly:1994ig}.)
Our main goal here is to investigate the validity of this equivalence also for all 
other $n$-photon amplitudes, which
have each a distinct dependence on $T$, in the HTL approximation.

%In order to obtain the HTL approximation it is convenient to perform the change of variable $|\vec k| = u T$ in Eq. \eqref{1pi1}.
%Then, 
Using the explicit expression for the
on-shell momentum $k_\mu = |\vec k| K_\mu$, with $K=(\sqrt{1 + m^2/|\vec k|^2},\vec k/|\vec k|)$, Eq. \eqref{1pi1}  yields
\begin{eqnarray}\label{forward2} 
{}^{\pm}\! \Pi^{htl} _{\mu_1\mu_2\dots\mu_n} &=& -  
%&=& xxxxx
\frac{1}{ (2\pi)^3}\int_0^\infty |\vec k|^2\, d|\vec k| \frac{N_{F}\left(\sqrt{(\vec k)^2+m^2} \right) }{2\sqrt{(\vec k)^2+m^2}}
\nonumber \\ &{} &
\int d \Omega \sum_{{perm.} \atop n}  {A}^n_{\mu_1\mu_2\dots\mu_n} (\pm |\vec k|  K,p_1,\cdots,p_n) .
\nonumber \\
\end{eqnarray}
%where $\int d \Omega$ denotes the integration over the directions of \hbox{$\hat K \equiv {\vec k}/{|\vec k|}$}.
%and $\beta\equiv 1/T$.
In the hard thermal loop approximation the denominators of the forward scattering amplitudes in \eqref{forward2}  
can be expanded using
\begin{equation}\label{htl1}
\frac{1}{s_n^2+2k\cdot s_n} = \frac{1}{|\vec k|} \frac{1}{2K\cdot s_n} - \frac{1}{|\vec k|^2} \frac{s_n^2}{(2 K\cdot s_n)^2} 
+ \dots .%;\;\; k = T u K.
\end{equation}
When combined with the momentum dependence of the numerators, as shown in Eq. \eqref{amp1}, the resulting expression
for the forward scattering amplitudes in Eq. \eqref{forward2} can be expressed as 
%turns out to be a sum of an infinite number of terms. 
\begin{widetext}
\begin{equation}\label{3.15}
 {A}^n_{\mu_1\mu_2\dots\mu_n}(\pm |\vec k| K,p_1,\cdots,p_n)  = \pm |\vec k| {A_1^n}_{\mu_1\mu_2\dots\mu_n} +  {A_2^n}_{\mu_1\mu_2\dots\mu_n} 
\pm \frac{1}{|\vec k|} {A_3^n}_{\mu_1\mu_2\dots\mu_n} + \frac{1}{|\vec k|^2}  {A_4^n}_{\mu_1\mu_2\dots\mu_n} 
\pm \dots      .
\end{equation}
\end{widetext}
%For all the contributions of order higher that ${A_3^n}_{\mu_1\mu_2\dots\mu_n}$, 
%it is understood that the $|\vec k|$ integration in Eq. \eqref{forward2} 
%has a infrared cut-off of order $m$ (the same prescription is to be understood for the amplitudes 
%with $n\ge 4$ obtained using the Boltzmann transport equation approach of the previous section). 
%A natural choice for $\mu$ is  the electron mass $m$ so that the higher point functions will  be
%expressed in terms of the ratio $\mu/T$. 
%In other words, we are not allowed to neglect the fermion mass for higher point functions.
%Nevertheless, the general structure of the integrands ${A}_{\mu_1\mu_2\mu_3\cdots A\mu_n}$ 
%is completely independent of  the regularization procedure and can be directly compared with the corresponding
%amplitudes in the Boltzmann transport equation approach of the previous section.
%Also, the temperature dependence of the finite part of these amplitudes can be obtained 
%in a systematic way as shown in the Appendix \ref{apb}.

Let us first consider the simplest case, namely the one-photon amplitude. In this case the forward scattering amplitude is exactly given by
%{\bf[ \color{red} sinal do loop fermionico e $i^2$ do vertice x propagador]}
\begin{equation}
{A}^1_{\mu}(\pm u T K)  = \mp e u T \,{\rm tr} \slashed{K} \gamma_\mu = \mp 4 e u T K_\mu ,
\end{equation}
where $u T \equiv |\vec k|$.
%and there are no permutations (of course there is also noexpansion in powers of $T$; the result is exact). 
Using Eq. \eqref{1pi1} we obtain the following result for the partial thermal one-photon amplitudes  
\begin{equation}
{}^{\pm}\! \Pi^{therm} _{\mu}  =    \pm
\frac{2 e T^3 I^{(1)}(\bar m)}{ (2\pi)^3}     %\int_0^\infty \frac{u^2\, du}{{\rm e}^u+1}
\int d \Omega K_\mu .
\end{equation}
Comparing this result with the left-hand side of Eq. \eqref{eqv1} (for $n=1$) , 
which can be easily obtained from Eq. \eqref{eq17}, we can see that 
Eq. \eqref{eqv1} is satisfied provided that ${\cal C}=2$, which is the correct result for fermions in $3+1$ dimensions.  

Next, let us consider the more interesting cases when $n\ge 2$.
As we have seen in the previous section, the Boltzmann transport equation approach produces $(n\ge 2)$-photon amplitudes which are 
proportional to $T^{4-n}\,I^{(n)}(\bar m)$. %(or logarithmic in $T$, in the case of the four-photon amplitude). 
On the other hand,  Eq. \eqref{3.15}  seems to yield, for each thermal amplitude, when substituted in Eq. \eqref{forward2},
several other terms proportional to higher powers of $T$.
%higher powers of $T$, starting with $T^3$. 
%, with the only exception of the one photon function, which has a single contribution proportional to $T^3$. 
Therefore, a necessary condition to have the equivalence \eqref{eqv1}, is that contributions proportional to higher 
powers of $T$ should vanish. For instance, the term proportional to $T^3 I^{(1)}(\bar m)$, in the two-photon function, 
as well as the terms proportional to $T^3 I^{(1)}(\bar m)$ and $T^2 I^{(2)}(\bar m)$ in the three-photon function, and so on, should vanish.

All the contributions proportional to $T^3 I^{(1)}(\bar m)$, which would contribute to \eqref{forward2},  
come from the first term in Eq. \eqref{3.15}. From the HTL procedure above described, one can see
that ${A^n_1}_{\mu_1\mu_2\dots\mu_n}$ must have the highest degree in $K$, being proportional to
\begin{equation}
\frac{K_{\alpha_1} K_{\alpha_2} \dots K_{\alpha_n}}{(K\cdot p_1) \, [K\cdot (p_1+p_2)]  \dots  [K\cdot (p_1+p_2+p_{n-1})]}  .
\end{equation}
After performing the sum over all the cyclic permutations and using momentum conservation, it is straightforward 
to show that the resulting expression vanishes.   More formally, we have the following necessary condition for the equivalence high-temperature QED and the Boltzmann transport equation formalism of the previous section.
\begin{equation}\label{eik1}
\sum_{\mbox{cycl.}}  {A^n_1}_{\mu_1\mu_2\dots\mu_n}(K,p_1,\cdots ,p_n) = 0 ;\;\;\; n\ge 2 .
\end{equation}
We remark that this is a property of individual diagrams which is true at the level of the forward scattering amplitudes (one does not need to add the $(n-1)!$ permutations to cancel these so-called superleading $T^3$ contributions). 
Of course for a neutral plasma the sum ${}^{+}\! \Pi^{htl} _{\mu_1\mu_2\dots\mu_n}   +  {}^{-}\! \Pi^{htl} _{\mu_1\mu_2\dots\mu_n}$ would vanish trivially,
since  the terms  which are proportional to any odd power of  $T$ are 
also odd in $k$. % (for the same reason the full amplitude in the case of odd $n$ should vanish in general accordingly to Furry's theorem).
However, in the present context, it is important to check the behavior of the 
individual components (for both even and odd $n$), in order to properly compare with the Boltzmann transport equation results of the previous section.

%This is indeed quite obvious, since in this case \eqref{eik1} reduces to 
%\begin{equation}
%\frac{1}{K\cdot p_1} + \frac{1}{K\cdot p_2}  = \frac{1}{K\cdot p_1} - \frac{1}{K\cdot p_1} = 0  .
%\end{equation}
%Similarly, in the case of the three-photon amplitude we would have
%\begin{widetext}
%\begin{equation}
%\frac{1}{K\cdot p_1 \, K\cdot p_2} +  \frac{1}{K\cdot p_2 \, K\cdot p_3} + \frac{1}{K\cdot p_3 \, K\cdot p_1}  =  
%\frac{K\cdot(p_1+p_2+p_3)}{K\cdot p_1 \, K\cdot p_2  \, K\cdot p_3} = 0.  
%\end{equation}
%\end{widetext}

Since Eq. \eqref{eik1}  already eliminates the $T^3$ contribution to the two-photon function, only the second
term in Eq. \eqref{3.15} may contribute to leading order, producing a $T^2$ contribution.
A straightforward calculation yields ($p_1=p$ and $p_2=-p$)
\begin{equation}\label{3.20}
{A_2^2}_{\mu_1\mu_2\dots\mu_n} = 2 e^2 \left( \eta_{\mu\nu} - \frac{K_\mu p_\nu + K_\nu p_\mu}{K\cdot p} + \frac{p^2 K_\mu K_\nu}{(K\cdot p)^2} \right).
\end{equation}
Inserting  Eq. \eqref{3.20} into Eq. \eqref{forward2}, we obtain
\begin{eqnarray}\label{twoTC} 
{}^{\pm}\! \Pi^{htl} _{\mu \nu} &=& -  
%&=&
\frac{2 e^2 T^2 I^{(2)}(\bar m)}{ (2\pi)^3}     %\int_0^\infty \frac{u\, du}{{\rm e}^u+1}
\nonumber \\ &{} & \!\!\!\!\!\!\!\!\!\!\!\!\!\!\!\!\!\!\!\!\!\!\!\!
\int d \Omega\left( \eta_{\mu\nu} - \frac{K_\mu p_\nu + K_\nu p_\mu}{K\cdot p} + \frac{p^2 K_\mu K_\nu}{(K\cdot p)^2} \right).
\end{eqnarray}
Comparing Eq. \eqref{twoTC}  with Eq. \eqref{eq33} we can see that \eqref{eqv1} is satisfied for ${\cal C} =2$, which is the correct result for fermions in $3+1$ dimensions.  

Up to this point the thermal field theory calculations have been performed without using any computer algebra tool. If we go further and try to verify 
Eq. \eqref{eqv1} for $n=3,4,\dots$  the calculation starts to become much more involved. Therefore, to proceed up to at least 
$n=4$ we have employed the FeynCalc computer algebra tool \cite{Mertig1991345}.

At the three-photon order, there would be in principle contributions from $A^3_2$, which 
would give rise to a contribution proportional to $T^2$. 
This would invalidate the equivalence of the Boltzmann transport equation approach with thermal field theory. 
However,  a  straightforward calculation yields
%\begin{widetext}
\begin{eqnarray}\label{e322}
{A_2^3}_{\mu_1\mu_2\mu_3} &=& \frac{2 e^3 K_{{\mu_3}}}{K\cdot {p_3}}\left(
 % -\frac{2 K_{{\mu_1}} g_{{\mu_2}{\mu_3}}}{K\cdot {p_1}}
  {\eta_{{\mu_1}{\mu_2}}} - \frac{K_{{\mu_1}} {p_1}_{{\mu_2}} + K_{{\mu_2}} {p_1}_{{\mu_1}}}{K\cdot {p_1} }
\right. \nonumber \\ 
 &+& \left.\frac{(p_1)^{2} K_{{\mu_1}} K_{{\mu_2}} }{(K\cdot {p_1})^{2}} \right)
 %-\frac{2 {p_3}^{2} K_{{\mu_1}} K_{{\mu_2}}K_{{\mu_3}}}{K\cdot {p_1} K\cdot {p_3}^{2}}
 %-\frac{2 K_{{\mu_2}} K_{{\mu_3}} {p_1}^{{\mu_1}}}{K\cdot {p_1} K\cdot{p_3}}
 %-\frac{2 K_{{\mu_1}} K_{{\mu_3}} {p_1}_{{\mu_2}}}{K\cdot {p_1} K\cdot {p_3}}
%\nonumber \\ &-&\left. 
- (\mu_1,p_1) \leftrightarrow (\mu_3,p_3)  , \nonumber \\
\end{eqnarray}
%\end{widetext}
which is manifestly antisymmetric. Therefore, the sum of its permutations vanish trivially.
(It is remarkable how  Eq. \eqref{e322} is simply related to the tensor structure of Eq. \eqref{twoTC}. Indeed,
if we contract it with  ${p_3}_{{\mu_3}}$ we obtain a relation which resembles the Lorentz structure of the non-Abelian gauge theory, without the antisymmetric colour factor).
%We will see that this property is explicitly verified for the case of four-photon function.

We are then left with a leading contribution to the three-photon function which is of order $T$ as in the case of the Boltzmann transport equation 
approach. With the help of computer algebra,
the resulting expression for ${A_3^3}_{\mu_1\mu_2\mu_3}$ can be computed in a straightforward way.
After performing the permutations of the external photon lines, we obtain complete agreement with
\eqref{eqv1}, when using Eqs. \eqref{eq38} and \eqref{a1}. 

The next and more interesting $n=4$-photon order can be dealt with similarly to the previous cases, although the 
calculation becomes much more involved. A straightforward computer algebra calculation shows that, 
besides the cancellation of the contribution proportional to $T^3$, the other contributions due 
to ${A^4_2}_{\mu_1\dots\mu_4}$ and ${A^4_3}_{\mu_1\dots\mu_4}$ also vanish after we add the permutations.
Finally, the remaining result is also in agreement with \eqref{eqv1} as in the previous cases.
This shows that, up to the four-photon order,  we have full agreement with the Boltzmann transport equation approach.

The previous results show that equivalence between thermal field theory and the Boltzmann transport equation approach 
is verified at the level of the forward scattering amplitudes, in the HTL limit, so that it is not necessary to explicitly 
perform the angular integrations  in Eq. \eqref{forward2}. 
%Let us now consider some simple examples when the integral can be explicitly performed in terms of elementary functions.
%
%\begin{equation}
%...
%\end{equation}
%
%{\bf \color{red} A ser continuado ... confiram tudo até aqui (inclusive o texto!)}

\section{Discussion}\label{sec4}
In this work we have investigated the equivalence between the high temperature limit of thermal field theory and the Boltzmann transport equation. 
Specifically, we have argued that both approaches produce the same $n$-photon thermal amplitudes. This is indicated by the general form derived
for the amplitudes, as well as from the explicit calculations up to $n=4$. Furthermore, we have considered the more general cases when the plasma may not
be neutral, so that the amplitudes with odd values of $n$ are nonvanishing. This equivalence implies that the quantities $A^n_p$ in Eq. \eqref{3.15}  
such that $p <  n$ should vanish. Only the contribution with $p=n$ can be interpreted as arising from the classical limit of finite temperature QED. On the other hand,  the $p > n$ contributions represent quantum corrections which are not described by the Boltzmann transport equation.
%xcvxcv

A similar equivalence is well known in the case of non-Abelian gauge theories as well as gravity 
and also in the particular case of the two-photon amplitude in QED. In the case of non-Abelian theories it has been
useful in order to derive a closed form for the gauge invariant hard thermal loop effective action \cite{kelly:1994ig}. 
Our analysis shows that all the results obtained from the Boltzmann transport equation can be identified with a more general HTL limit of the thermal Green's functions in QED.  Consequently, the gauge invariant effective action, rather than being proportional to a single power of the temperature,  
will have a more general dependence on $T$ as can be seen from the form of the thermal amplitudes in Eq. \eqref{eq38a}. 
We remark that the electron mass is not being neglected so that the amplitudes with $n\ge 4$ are well defined functions of $m/T$.

Once the equivalence with the Boltzmann transport equation is established, a full resumation of the thermal effects in QED, to all orders in the external field, would require a further investigation, taking into account the gauge invariant effective actions as given by Eq. \eqref{eq26}.
% on the possibility of summing to all orders in the external field.
The resulting effective action would be relevant to describe a low density plasma in an external field with frequencies which are small compared with the temperature.

\acknowledgments
We would like to thank CNPq and CAPES (Brazil) for financial support.

%\newpage

\appendix

\section{}\label{apa}
Here we present the result for the integrand of the three-photon amplitude \eqref{eq38}:
\begin{widetext}
 \begin{eqnarray}\label{a1}
{\cal A}_{\mu_1\mu_2\mu_3}  &=& \sum_{{perm.} \atop 3} \left[
-{\frac {{ K_{\mu_3}}\,{ K_{\mu_2}}\,{ K_{\mu_1}}\,{ p_2\cdot {p_3}}\,{ {p_1}^2}}{{{ (K\cdot p_3)}}^{2}{{ (K\cdot p_1)}}^{2}}}-
{\frac {2\,{ K_{\mu_3}}\,{ K_{\mu_2}}\,{ K_{\mu_1}}\,{ p_1\cdot p_3}\,{ p_2\cdot {p_3}}}{{{ (K\cdot p_3)}}^{3}{ K\cdot p_1}}}+
{\frac {{ K_{\mu_3}}\,{ K_{\mu_2}}\,{ {p_1}_{\mu_1}}\,{ p_2\cdot {p_3}}}{{{ (K\cdot p_3)}}^{2}{ K\cdot p_1}}}+
\right.\nonumber \\ &&
{\frac {2\,{ K_{\mu_3}}\,{ K_{\mu_2}}\,{ {p_3}_{\mu_1}}\,{ p_2\cdot {p_3}}}{{{ (K\cdot p_3)}}^{3}}}+ 
{\frac {{ K_{\mu_3}}\,{ K_{\mu_1}}\,{ {p_3}_{\mu_2}} \,{ K\cdot p_2}\,{ {p_1}^2}}{{{ (K\cdot p_3)}}^{2}{{ (K\cdot p_1)}}^{2}}}-
{\frac {{ K_{\mu_3}}\,{ K_{\mu_1}}\,{ {p_3}_{\mu_2}}\,{ p_1\cdot p_2}}{{{ (K\cdot p_3)}}^{2}{ K\cdot p_1}}}+
{\frac {{ K_{\mu_3}}\,{ K_{\mu_1}}\,{ {p_1}_{\mu_2}}\,{ p_2\cdot {p_3}}}{{{ (K\cdot p_3)}}^{2}{ K\cdot p_1}}}+
\nonumber \\ &&
{\frac {2\,{ K_{\mu_3}}\,{ K_{\mu_1}}\,{ {p_3}_{\mu_2}}\,{ K\cdot p_2}\,{ p_1\cdot p_3}}{{{ (K\cdot p_3)}}^{3}{ K\cdot p_1}}}- 
{\frac {{ K_{\mu_3}}\,{ {{p_1}}_{\mu_1}}\,{ {p_3}_{\mu_2}} \,{ K\cdot p_2}}{{{ (K\cdot p_3)}}^{2}{ K\cdot p_1}}}+
{\frac {{ K_{\mu_3}}\,{ {p_2}_{\mu_1}}\,{ {p_3}_{\mu_2}}}{{{ (K\cdot p_3)}}^{2}}}-
{\frac {{ K_{\mu_3}}\,{ \eta_{\mu_1\mu_2}} \,{ p_2\cdot {p_3}}}{{{ K\cdot p_3}}^{2}}}-
\nonumber \\ &&
{\frac {2\,{ K_{\mu_3}}\,{ {p_3}_{\mu_1}}\,{ {p_3}_{\mu_2}} \,{ K\cdot p_2}}{{{ (K\cdot p_3)}}^{3}}}+
{\frac {{ K_{\mu_2}}\,{ K_{\mu_1}}\,{ {p_2}_{\mu_3}}\,{ {p_1}^2}}{{ K\cdot p_3}\,{{ (K\cdot p_1)}}^{2}}}+
{\frac {{ K_{\mu_2}}\,{ K_{\mu_1}}\,{ {p_2}_{\mu_3}}\,{ p_1\cdot p_3}}{{{ (K\cdot p_3)}}^{2}{ K\cdot p_1}}}+
{\frac {{ K_{\mu_2}}\,{ K_{\mu_1}}\,{ {p_1}_{\mu_3}}\,{ p_2\cdot {p_3}}}{{{ (K\cdot p_3)}}^{2}{ K\cdot p_1}}}-
\nonumber \\ &&
{\frac {{ K_{\mu_2}}\,{ {p_1}_{\mu_1}}\,{ {p_2}_{\mu_3}}}{{ K\cdot p_3}\,{ K\cdot p_1}}}- 
{\frac {{ K_{\mu_2}}\,{ {p_3}_{\mu_1}}\,{ {p_2}_{\mu_3}}}{{{ (K\cdot p_3)}}^{2}}}-
{\frac {{ K_{\mu_2}}\,{ p_2\cdot {p_3}}\,{ \eta_{\mu_1\mu_3}}}{{{ (K\cdot p_3)}}^{2}}}-
{\frac {{ K_{\mu_1}}\,{ K\cdot p_2}\,{ {p_1}^2}\,{ \eta_{\mu_2\mu_3}}}{{ K\cdot p_3}\,{{ (K\cdot p_1)}}^{2}}}-
{\frac {{ K_{\mu_1}}\,{ {p_2}_{\mu_3}}\,{ {p_1}_{\mu_2}}}{{ K\cdot p_3}\,{ K\cdot p_1}}}+\nonumber \\  &&
{\frac {{ K_{\mu_1}}\,{ p_1\cdot p_2}\,{ \eta_{\mu_2\mu_3}}}{{ K\cdot p_3}\,{ K\cdot p_1}}}-
{\frac {{ K_{\mu_1}}\,{ K\cdot p_2}\,{ {p_3}_{\mu_2}}\,{ {p_1}_{\mu_3}}}{{{ (K\cdot p_3)}}^{2}{ K\cdot p_1}}}-
{\frac {{ K_{\mu_1}}\,{ K\cdot p_2}\,{ p_1\cdot p_3}\,{ \eta_{\mu_2\mu_3}}}{{{ (K\cdot p_3)}}^{2}{ K\cdot p_1}}}+
{\frac {{ K\cdot p_2}\,{ {p_1}_{\mu_1}}\,{ \eta_{\mu_2\mu_3}}}{{ K\cdot p_3}\,{ K\cdot p_1}}}- .\nonumber \\  &&
\left.
{\frac {{ {p_2}_{\mu_1}}\,{ \eta_{\mu_2\mu_3}}}{{ K\cdot p_3}}}+{\frac {{ {p_2}_{\mu_3}}\,{ \eta_{\mu_1\mu_2}}}{{ K\cdot p_3}}}+
{\frac {{ K\cdot p_2}\,{ {p_3}_{\mu_1}}\,{ \eta_{\mu_2\mu_3}}}{{{ (K\cdot p_3)}}^{2}}}+
{\frac {{ K\cdot p_2}\,{ {p_3}_{\mu_2}}\,{ \eta_{\mu_1\mu_3}}}{{{ (K\cdot p_3)}}^{2}}} \right]
\end{eqnarray}
\end{widetext}
A similar but much more involved expression has been obtained for the four-photon amplitude, using computer algebra. 
In all the cases we have obtained agreement between the results derived using either the Boltzmann transport equation or the high temperature limit of thermal field theory.

\end{document}